\documentclass[review]{elsarticle}

\usepackage{hyperref}

\journal{Elsevier Journal of Ocean Engineering}

\DeclareGraphicsExtensions{.eps}
\usepackage{array}
\usepackage{multirow}
\usepackage{hhline}
\usepackage{amssymb, latexsym}
\usepackage{amsmath}
\usepackage{amsthm}
\usepackage{enumerate}
\usepackage{subfig}
\usepackage{color}

\makeatletter
\newcommand{\Thickhline}{%
    \noalign {\ifnum 0=`}\fi \hrule height 1pt
    \futurelet \reserved@a \@xhline
}
\def\M{\mathrm{NMF}}
\newcolumntype{P}[1]{>{\centering\arraybackslash}p{#1}}
\newcolumntype{'}{!{\vrule width 2pt}}
\def\ve#1{{\mathchoice{\mbox{\boldmath$\displaystyle #1$}}%
              {\mbox{\boldmath$\textstyle #1$}}%
              {\mbox{\boldmath$\scriptstyle #1$}}%
              {\mbox{\boldmath$\scriptscriptstyle #1$}}}}
\makeatother

\begin{document}

\begin{frontmatter}

\title{Computationally Efficient Calculations of Target Performance of the Normalized Matched Filter Detector for Hydrocoustic Signals}

\author{Roee Diamant\\
Department of Marine Technology, The University of Haifa\\
Email: roeed@univ.haifa.ac.il}

%

\begin{abstract}
Detection of hydroacoustic transmissions is a key enabling technology in applications such as depth measurements, detection of objects, and undersea mapping. To cope with the long channel delay spread and the low signal-to-noise ratio, hydroacoustic signals are constructed with a large time-bandwidth product, $N$. A promising detector for hydroacoustic signals is the normalized matched filter (NMF). For the NMF, the detection threshold depends only on $N$, thereby obviating the need to estimate the characteristics of the sea ambient noise which are time-varying and hard to estimate. While previous works analyzed the characteristics of the normalized matched filter (NMF), for hydroacoustic signals with large $N$ values the expressions available are computationally complicated to evaluate. Specifically for hydroacoustic signals of large $N$ values, this paper presents approximations for the probability distribution of the NMF. These approximations are  found extremely accurate in numerical simulations. We also outline a computationally efficient method to calculate the receiver operating characteristic (ROC) which is required to determine the detection threshold. Results from an experiment conducted in the Mediterranean sea at depth of 900~m agree with the analysis.
\end{abstract}


\end{frontmatter}


\section{Introduction}\label{sec:intro}

Underwater acoustics can fulfil the needs of a multitude of underwater applications. This include: oceanographic data collection, warning systems for natural disasters (e.g., seismic and tsunami monitoring), ecological applications (e.g., pollution, water quality and biological monitoring), military underwater surveillance, assisted navigation, industrial applications (offshore exploration), to name just a few \cite{Akyildiz:2006}. Detection of hydroacoustic signals is characterized by a target probability of false alarm and probability of detection. The detection is performed for a buffer of samples, $y(t)$, recorded from the channel (usually in a sliding time window fashion). In this paper, the focus is on detection of signals of known structure. The applications in mind are active sonar systems, acoustic localization systems (e.g., ultra-short baseline), and acoustic systems used for depth estimation, ranging, detection of objects, and communications.

In this paper, we focus on the first step in the detection chain, namely, a binary hypothesis problem where the decoder differentiate between a \textit{noise-only} hypothesis and a \textit{signal exist} hypothesis. The former is when the sample buffer, $y(t)$, consists of ambient noise, and the latter is the case where the sample buffer also includes a distinct received hydroacoustic signal.  Without channel state information, the most common detection scheme is the matched filter \cite{Burdic:2002}, which is optimal in terms of the signal-to-noise ratio (SNR) in case on an additive white Gaussian channel. The matched filter detector is a constant false alarm rate (CFAR) test, and its detection threshold is determined only by the target false alarm probability (cf. \cite{kazakos:1990}). Due to the (possibly) large dynamic range of the detected signal \cite{Bumiller:2007}, and for reasons of template matching \cite{Conte:1998}, the matched filter is often normalized by the noise covariance matrix. This normalization is often referred to as adaptive normalized matched filter (ANMF) and is the preferred choice in several tracking applications such as gradient descent search, active contour models, and wavelet convolution \cite{Lewis:1995}.

To estimate the noise covariance matrix, several noise-only training signals are required \cite{Kelly:1989}. Since this limits the application, and since the noise may be time-varying, various ANMF detectors have been developed. Based on the noise texture model, \cite{Younsi:2011} suggested a maximum likelihood estimator for the noise covariance matrix. Alternatively, in \cite{Robey:1992} an iterative procedure is performed where first the covariance matrix is assumed known and the test statistics for a signal vector is calculated. Next, using these statistics and additional noise-only vectors, the noise covariance matrix is estimated and is substituted back into the test statistics. In \cite{Kraut:2001}, an adaptive matched subspace detector is developed and its statistical behavior is analyzed to adapt the detector to unknown noise covariance matrices in cases where the received signal is distorted compared to transmitted one.

The above normalization methods of the matched filter require an estimation of the covariance matrix of the ambient noise. As shown in \cite{Scharf:1971}, mismatch in this estimation effects detection performance and target false alarm and detection rates may not be satisfied. Since in underwater acoustics the noise characteristics are often fast time varying \cite{Burdic:2002}, an alternative detection scheme is to normalize the matched filter with the power of $y(t)$ \cite{Scharf:2000}, \cite{Burdic:2002}. We refer to this scheme as the \textit{normalized matched filter} (NMF), as opposed to the ANMF. The NMF detector does not require estimation of the noise covariance matrix. Instead, its detection threshold depends only on the time-bandwidth product, $N$, of the expected signal. For underwater applications which require detection at target performance in various noise conditions, the NMF may be a suitable choice.

In \cite{Rangaswamy:2002}, a low-rank NMF is suggested, where the linear matched filter is normalized by the power of the transmitted signal and a projection of the detected one. The projection is made according to the estimated noise covariance matrix, and the result is a simplified test which is proportional to the output of the standard colored-noise matched filter. A modification of the matched filter is proposed in \cite{Bumiller:2007} for the case of a multipath channel. The works in \cite{Bumiller:2007} and \cite{Rangaswamy:2002} include analysis for the false alarm and detection probabilities of the NMF. This analysis is either a modification of a similar study of the NMF or is based on semi-analytic matrix representation. 

Due to low signal-to-noise ratio and the existence of narrow band interferences, hydroacoustic signals are constructed with a large time-bandwidth product of typical values $N>50$ \cite{Mason:2008, Rouseff:2009, Walree:2013}. While the NMF has been analyzed before, for large $N$ the available expressions are computationally complicated to evaluate. Consequently, it is difficult to evaluate the receiver operating characteristic (ROC), which is required to determine the detection threshold. As a result, most underwater applications avoid using the NMF as a detector. Considering this problem and based on the probability distribution of the NMF and its moments, in this paper computationally efficient approximations for the probability of false-alarm and for the probability of detection for signals of large $N$ are offered. This leads to a practical scheme for the evaluation of the ROC. Simulation results show that the developed expressions are extremely accurate in the large $N$ limit. To test the correctness of the analysis in real environment, results from a sea experiment are reported. The experiment was conducted in the Mediterranean sea to detect chirp signals reflected from the sea bottom at depth of 900~m.

The reminder of this paper is organized as follows. The system model is presented in Section~\ref{sec:model}. In Section~\ref{sec:distribution}, we derive the probability distribution of the NMF and give expressions for the probability of false alarm and for the probability of detection. Next, performance evaluation in numerical simulation (Section~\ref{sec:simulation}) and results from the sea experiment (Section~\ref{sec:experiement}) are presented in Section~\ref{sec:performance}. Finally, conclusions are drawn in Section~\ref{sec:Conclusions}. The notations used in this paper are summarized in Table~\ref{l:notation}.

\begin{table}[ht]
\begin{tabular}{ll}
\hline\hline
Notation & Explanation \\ [0.5ex]
\hline
$y(t)$ & Received data from the channel\\
$s(t),s_k$ & transmitted signal and its $k$th sample, respectively\\
$n(t),n_k$ & Channel ambient noise and its $k$th sample, respectively\\
$\sigma^2$ & variance of channel ambient noise\\
$T$ & duration of signal\\
$W$ & Bandwidth of signal\\
$N$ & product of bandwidth and duration of signal \\
$\M$ & output of normalised matched filter\\
$x=\cos(\theta_{N-2})$ & Output of NMF for $y(t)=n(t)$\\
$x_T=\cos(\theta_{T})$ & Detection threshold\\
$\cos(\phi)$ & Output of NMF for $y(t)=s(t)+n(t)$\\
$P_{\mathrm{FA}}$ & probability of false alarm\\
$P_{\mathrm{D}}$ & probability of detection\\
[1ex]
\hline
\end{tabular}
\caption{List of major notations}
\label{l:notation}
\end{table}

\section{System Model}\label{sec:model}

The goal of this paper is to offer a computational efficient determination of the detection threshold of the NMF for signals with large $N$ property. Since the NMF is executed at the very first step of the reception chain,  the receiver poses no information of the channel or range to transmitter. Therefore, only an additive noise of unknown variance can be assumed for the system model. 

For a received signal, $y(t)$, we consider a binary detection test of hypotheses,
\begin{eqnarray}
&H_0&: \ y(t)=n(t)\;,\nonumber\\
&H_1&: \ y(t)=s(t)+n(t)\;.
\label{e:hypo}
\end{eqnarray}
In (\ref{e:hypo}), $s(t)$ is an hydroacoustic signal of bandwidth $W$, duration $T$, and $n(t)$ is an additive noise. Let us define the time-bandwidth product $N=WT$. We assume that $N$ is large (values exceeding 50 are enough). In our analysis we consider the case of real signals. However, as demonstrated in Section~\ref{sec:performance}, the analysis holds for the case of complex signals.

We are interested in the following quantity (referred to as the NMF),
\begin{eqnarray}
\M&=&\frac{\left|\int s(t)y(t)dt\right|}{\sqrt{\int s^2(t)dt\int y^2(t)dt}}\nonumber\\
&=&\frac{\left|\sum\limits_{k=1}^N s_ky_k\right|}{\sqrt{\sum\limits_{k}s_k^2\sum\limits_{l}y_l^2}}\;,
\label{e:NMF}
\end{eqnarray}
where $s_k$ and $y_k$ are the $k$th sample of $s(t)$ and $y(t)$, respectively, and $y(t)$ is sampled equally at the Nyquist rate. For a detection scheme which uses correlator (\ref{e:NMF}) as its detection metric, the objective is to develop computational efficient expression for the probability of false alarm and for the probability of detection. Both figures are required to determine the detection threshold through the ROC.

The strong assumption in this paper is of i.i.d zero-mean Gaussian noise $n(t)$ with variance $\sigma^2$. As discussed in Section~\ref{sec:simulation}, effect of mismatch in the noise model is shown negligible. However, the case of coloured noise can be treated by including a trivial whitening mechanism in the filtering process. Namely (\ref{e:NMF}) becomes,
\begin{equation}
\frac{\sum\limits_{j,k}^N s_jw_{j,k}y_k}{\sqrt{\sum\limits_{j,k}s_jw_{j,k}s_k\sum\limits_{j',k'}y_{j'}w_{j',k'}y_{k'}}}\;,
\label{e:NMF2}
\end{equation}
where $w$ is the inverse correlation-matrix satisfying $\sum\limits_{k}w_{j,k}E\left[n_k,n_l\right]=\delta_{j,k}$ and $\delta$ is the Kronecker delta function. The following results can therefore be generalized without the need of significant modifications.

\section{Probability Distribution Analysis}\label{sec:distribution}

In this section, we formulate the probability distribution of the NMF and for large $N$, give approximations for the probability of false alarm and for the probability of detection.

\subsection{Probability of False Alarm}\label{sec:pfa}

\begin{figure}[t]
\centering
\includegraphics[width=4.0in]{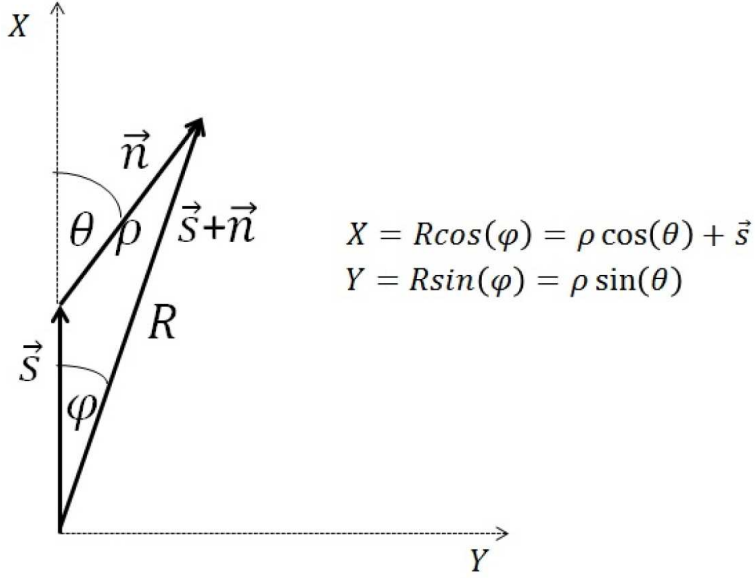}
\caption{Spherical coordinates of received signal $s(t)$ and noise $n(t)$.}
\label{f:1}
\end{figure} 
Let $\vec{\ve{s}}$, $\vec{\ve{n}}$ be N-dimensional space vectors whose elements are $s_k$ and $n_k$, respectively. It is easier to manage the following analysis using spherical coordinates. To this end, we set $\vec{\ve{s}}$ along the polar-axis (see Fig.~\ref{f:1}), such that $\rho^2=\sum\limits_{k}n_k^2$. The assumption of i.i.d Gaussian noise leads to the probability density function
\begin{equation}
P\left(\rho,\phi,\theta_1,\ldots,\theta_{N-2}\right)\partial\rho\partial\phi\prod\limits_{k=1}^{N-2}\partial\theta_k=\left(2\pi\sigma^2\right)^{-\frac{N}{2}}e^{-\frac{1}{2\sigma^2}\sum\limits_{i}n_i^2}\prod\limits_{l}\partial n_l\;.
\label{e:distibute}
\end{equation}
Then, for a noise-only signal, i.e., $y(t)=n(t)$, the NMF is given by the angle $\theta_{N-2}$ between vectors $\vec{\ve{s}}$ and $\vec{\ve{n}}$, such that
\begin{equation}
\M=\frac{\vec{s}\cdot\vec{n}}{|\vec{s}||\vec{n}|}=\cos{\theta_{N-2}}\;.
\label{e:NMF3}
\end{equation}
To find the probability of false alarm, we first need to evaluate the distribution $P(\theta_{N-2})$. Then, given a detection threshold $x_T$, we obtain
\begin{equation}
\hat{P}_{\mathrm{fa}}=\int\limits_0^{x_T}P(\theta_{N-2})d\theta_{N-2}\;.
\label{e:pfa_general}
\end{equation}

Let the volume-element, $dV=\prod\limits_{l}\partial n_l$, be expressed in terms of the solid angle $d\Omega$ such that for $0\leq\phi\leq2\pi,0\leq\theta_k\leq\pi$,
\begin{eqnarray}
dV&=&\rho^{N-1}d\rho d\Omega\nonumber\\
&=&\rho^{N-1}\partial\rho\partial\phi\prod\limits_{k=1}^{N-2}\partial\theta_k\sin^k\left(\theta_k\right)\;.
\label{e:omega}
\end{eqnarray}
Then, by integrating (\ref{e:distibute}) over all angular variables, except for the polar-angle $\theta_{N-2}$, one immediately obtains
\begin{equation}
P(\rho,\theta_{N-2})\approx C_{N,1}\rho^{N-1}e^{-\frac{\rho^2}{2\sigma^2}}\sin^{N-2}(\theta_{N-2}), \ 0\leq\theta_{N-2}\leq\pi,0\leq\rho<\infty\;,
\label{e:distibute2}
\end{equation}
where $C_{N,1}$ is a constant. Further integration over $\rho$ leads to
\begin{equation}
P(\theta_{N-2})=C_{N,2}\sin^{N-2}(\theta_{N-2})\;,\label{e:distibute3_a}
\end{equation}
and $C_{N,2}$ is a constant. 

For convenience, denote $x=\cos(\theta_{N-2})$. Expression (\ref{e:distibute3_a}) implies that all the odd moments of $x$ vanish identically, whereas even moments are given by
\begin{equation}
E\left[x^{2p}\right]=\frac{\Gamma\left(\frac{N}{2}\right)}{\sqrt(\pi)}\frac{\Gamma\left(p+\frac{1}{2}\right)}{\Gamma\left(p+\frac{N}{2}\right)}, \ p=0,1,2,\ldots\;.
\label{e:moment}
\end{equation}
In particular,
\begin{equation}
E\left[x^{2}\right]=\frac{1}{N}\;.
\label{e:moment2}
\end{equation}
The result in (\ref{e:moment2}) can be obtained directly by the method described in \ref{sec:appendix}, which confirms the above analysis. 

By (\ref{e:distibute3_a}) and (\ref{e:moment2}), when $N>>1$ the distribution $P(\theta_{N-2})$ approaches the Gaussian limit with the variance being $\frac{1}{N}$. Then, the probability of false alarm is approximated by
\begin{equation}
\hat{P}_{\mathrm{fa}}=\frac{1}{2}\mathrm{erfc}\left(x_T\sqrt{\frac{N}{2}}\right)\;,
\label{e:pfa2}
\end{equation}
However, since usually $P_{\mathrm{fa}}<<1$, unless $N$ is huge such that $P_{\mathrm{fa}}N>>1$ expression (\ref{e:pfa2}) is not accurate enough. Instead, the accurate term for the probability of false alarm is
\begin{equation}
P_{\mathrm{fa}}=1-B\left(x_T^2,\frac{1}{2},\frac{N-1}{2}\right)\;,
\label{e:pfa}
\end{equation}
where
\[
 B(a,b,z)=\int\limits_0^a t^{b-1}(1-t)^{z-1}dt
 \]
denotes the (tabulated) regularized incomplete beta function. Note that (\ref{e:pfa}) does not require calculation of the noise characteristics. 

\subsection{Probability of Detection}\label{sec:pd}

\subsubsection{Exact Term}

Suppose $y(t)=s(t)+n(t)$, and mark $s^2$ as the energy of the received signal. Setting $\vec{\ve{s}}$ along the polar-axis (see Fig.~\ref{f:1}) we have $\vec{\ve{y}}\cdot\vec{\ve{s}}=R\cos(\phi)$ with $\M=\cos(\phi)$. Therefore, changing variables $(\rho,\theta_{N-2})$ into $(R,\phi)$ in (\ref{e:distibute2}) we obtain
\begin{equation}
P(R,\phi)\approx R^{N-1}\sin^{N-2}(\phi)e^{-\frac{(R\cos(\phi)-s)^2+R^2\sin^2(\phi)}{2\sigma^2}}, \ -\leq\phi\leq\pi,0\leq R<\infty\;.
\label{e:pd_distribution}
\end{equation}
Integrating over $R$, $P(\phi)$ can be written in terms of the parabolic-cylinder function, 
\[
D_p(z)=\frac{1}{\pi}\int\limits_0^{\pi}\sin\left(p\alpha)-z\sin(\alpha)\right)d\alpha\;,
\]
i.e.,
\begin{equation}
P(\phi)=\pi^{-\frac{1}{2}}2^{1-\frac{N}{2}}\frac{\Gamma(N)}{\gamma\left(\frac{N-1}{2}\right)}e^{-s^2\frac{\frac{1}{2}-\frac{1}{4}\cos^2(\phi)}{\sigma^2}}\sin^{N-2}(\phi)D_{N}\left(-s\frac{\cos(\phi)}{\sigma}\right)\;.
\label{e:pd_distribution2}
\end{equation}
Alternatively, by the definition of $D_p(z)$,
\begin{equation}
\begin{split}
P(\phi)=\frac{e^{-\frac{s^2}{2\sigma^2}}\sin^{N-2}(\phi)}{\sqrt{\pi}\Gamma\left(\frac{N-1}{2}\right)}\cdot\left[\Gamma\left(\frac{N}{2}\right)F\left(\frac{N}{2},\frac{1}{2},\frac{s^2\cos^2(\phi)}{2\sigma^2}\right)+\right. \\
\left.\Gamma\left(\frac{N+1}{2}\right)\sqrt{2}s\frac{\cos(\phi)}{\sigma}F\left(\frac{N+1}{2},\frac{3}{2},\frac{s^2\cos^2(\phi)}{2\sigma^2}\right)
\right]\;,
\end{split}
\label{e:pd_distribution3}
\end{equation}
where 
\[
F(a,b,z)=\frac{\Gamma(b)}{\Gamma(b-a)\Gamma(a)}\int\limits_0^1 e^{zt}t^{a-1}(1-t)^{b-a-1}dt
\] 
is the confluent hypergeometric function. Note that as $\frac{s}{\sigma}\rightarrow0$, (\ref{e:pd_distribution3}) is reduced back to (\ref{e:distibute3_a}). The average NMF, derived from (\ref{e:pd_distribution3}), is given by the Kummer function,
\begin{equation}
E\left[\cos(\phi)\right]=\frac{\Gamma\left(\frac{N+1}{2}\right)}{\Gamma\left(\frac{N+2}{2}\right)}\frac{se^{-\frac{s^2}{2\sigma^2}}}{\sqrt{2\sigma^2}}F\left(\frac{N+1}{2},\frac{N+2}{2},\frac{s^2}{2\sigma^2}\right)\;.
\label{e:MF_average}
\end{equation}

The probability of detection for the detection threshold, $x_T$, can be found by
\begin{equation}
P_D=\int_{0}^{x_T}P(\phi)d\phi\;.
\label{e:pd}
\end{equation}
Unfortunately, for large $N$ direct numerical calculation of $P_D$ is bound to fail. This is because $P(\phi)$ contains infinitely many terms which oscillate rapidly as $N>>1$. It is therefore important to obtain asymptotic expressions for $P(\phi)$ in the large-$N$ limit.

\subsubsection{Approximated Solution}

When both $N$ and $\frac{s}{\sigma}$ are large compared to unity, $P(\phi)$ can be approximated using the asymptotic form of $D_p(z)$ \cite{Gradshteyn:2004},
\begin{equation}
D_p(z)\approx e^{-\frac{z^2}{4}}z^p\left(1+{\cal O}\left(\frac{z}{p}\right)\right)\;,
\label{e:pd_approx}
\end{equation}
applicable for $z>>1$ and $|z|>>|p|$ (i.e., for large SNR). However, this may not be applicable to all considered cases. Instead, the \textit{correct} asymptotic can be found by expanding $P(\phi)$ around its saddle-point.

To that end, let us go back to expression (\ref{e:pd_distribution}). Denoting $R\rightarrow \tilde{R}\frac{\sqrt{N}}{\sigma}$, and introducing $\gamma=\frac{s}{2\sigma\sqrt{N}}$, (\ref{e:pd_distribution}) takes the form
\begin{equation}
P_{\gamma}(\tilde{R},\phi)\approx\left(\tilde{R}\sin^2(\phi)\right)^{-1}e^{Ng}\;,
\label{e:pd_distribution5}
\end{equation}
where $g=\ln(\tilde{R})+\ln(\sin(\phi))-\frac{1}{2}\tilde{R}^2+2\tilde{R}\gamma\cos(\phi)$. Note that $\gamma$ is a function of $\frac{s}{\sigma}$ (which corresponds to the SNR). Since $P_{\gamma}=0$ at the end points $(\tilde{R},\phi)=(0,0)$ and $(\tilde{R},\phi)=(\infty,\pi)$, the large-$N$ behaviour of this function is dominated by Gaussian fluctuations around some saddle-points in the complex $(\tilde{R}\times\phi)$-hyper plane. The saddle-points equations are then
\begin{eqnarray}
\frac{\partial g}{\partial\tilde{R}}&=&\tilde{R}^{-1}-\tilde{R}+2\gamma\cos(\phi)=0\;,\label{e:saddle_a}\nonumber\\
\frac{\partial g}{\partial\phi}&=&\cot(\phi)-2\tilde{R}\gamma\sin(\phi)=0\;,
\label{e:saddle_b}
\end{eqnarray}
with fluctuations determined by the following Hessian (also known by the name ''Fisher information matrix'')
\begin{equation}
H=\left(\begin{array}{cc}\frac{\partial^2g}{\partial\tilde{R}^2} & \frac{\partial^2g}{\partial\tilde{R}\partial\phi} \\ \frac{\partial^2g}{\partial\phi\partial\tilde{R}} & \frac{\partial^2g}{\partial\phi^2}\end{array}\right)=-\left(\begin{array}{cc}\tilde{R}^{-2}+1 & 2\gamma\sin(\phi) \\ 2\gamma\sin(\phi) & 2\tilde{R}\gamma\cos(\phi+\frac{cos(\phi)}{\sin^2(\phi)})\end{array}\right)\;.
\label{e:hassian}
\end{equation}
Equation (\ref{e:saddle_b}) is solved by the quartet
\begin{eqnarray}
R_c&=&\frac{2\gamma^2\pm\sqrt{4\gamma^4+4\gamma^2+1}}{\sqrt{4\gamma^2+1}}\;,\\
\sin(\phi_c)&=&\pm\frac{1}{\sqrt{4\gamma^2+1}}\;.
\label{e:hassian2}
\end{eqnarray}
Fortunately, only one of these solutions (the one for which $R_c,\phi_c\geq0$) is reachable by a continuous deformation of the contour of integration. Substituting back into (\ref{e:hassian}), one obtains
\begin{equation}
|g''|=-\frac{\partial^2g}{\partial\phi_c^2}=1+4\gamma^2+\frac{4\gamma^4}{1+4\gamma^2}\left(2\gamma^2+\sqrt{4\gamma^4+4\gamma^2+1}\right)\;.
\label{e:hassian3}
\end{equation}
To the leading order in powers of $N^{-1}$ and for arbitrary values of $\gamma\geq0$,
\begin{equation}
P_{\gamma}(\phi)\approx\sqrt{\frac{N|g''|}{2\pi}}e^{-\frac{N}{2}|g''|(\phi-\phi_c)^2}, \ N>>1\;.
\label{e:pd_distribution4}
\end{equation}

For $\gamma<<1$ (i.e., small SNR), we get $\phi_c\approx\left(\frac{\pi}{2}-2\gamma\right)$ and $|g''|\approx1$. Therefore, (\ref{e:pd_distribution4}) implies that the NMF maintains good deflection as long as $\gamma N>1$, which is similar to other compressing filters. (Note that under this condition, the variance of $\phi$ is smaller than the SNR separation). In the opposite limit, as $\gamma$ increases, $\phi_c\approx2\gamma^{-1}$ approaches towards the edge-point $\phi=0$. At the same time, however, $|g''|\rightarrow4\gamma^4$ and $\frac{\mathrm{var}(\phi)}{\phi_c^2}\approx\frac{N\gamma^2}{-1}<<1$. Thus, when $\phi_c\rightarrow0$, the Gaussian lube shrinks thereby avoiding any significant deformations due to edge-effects. As a result the probability of detection, $P_d$, can be evaluated as
\begin{equation}
P_D=\frac{1}{2}\mathrm{erfc}\left((\phi_c-\theta_T)\sqrt{\frac{N|g''|}{2}}\right), \ \phi_c<\theta_T<\frac{\pi}{2}\;.
\label{e:pd2}
\end{equation}
This approximation introduces a relative error of the order ${\cal O}\left(N^{-1}\right)$ in the estimation of $P_D$. It follows from (\ref{e:pd2}) that, for a fixed $\frac{s}{\sigma}$, as the number of samples $N$ is increased, $P_d$ is saturated.

\begin{figure}[t]
\centering
\includegraphics[width=4.0in]{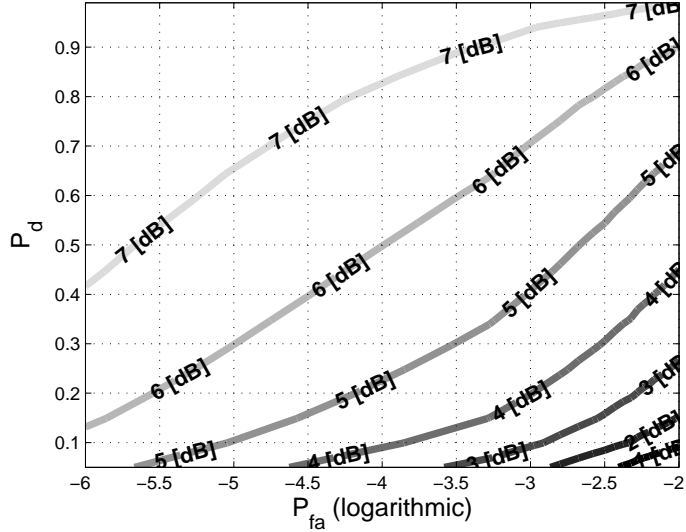}
\caption{ROC curves for $N=100$. Contour lines represents $s$ / $\sigma$ values. Logarithm base 10.}
\label{f:2}
\end{figure} 
Having expressions (\ref{e:pfa2}), (\ref{e:hassian3}), and (\ref{e:pd2}), one can construct the ROC in the large-$N$ limit. First, the detection threshold is obtained by inverting (\ref{e:pfa2}). Next, $|g''|$ is calculated with the help of (\ref{e:pd2}). Finally, the required $s/\sigma$ ratio is determined by solving (\ref{e:hassian3}) for $\gamma$. The resulting ROC curves for $N=100$ are shown in Fig.~\ref{f:2}.

\section{Performance Evaluation}\label{sec:performance}

To evaluate the accuracy of the expressions for $P_{\mathrm{fa}}$ and $P_D$, results from numerical simulations and from a sea experiment are now presented. To that end, the above analysis is compared with empirical measurements of the probability of false alarm, $\hat{P}_{\mathrm{fa}}$, and the probability of detection, $\hat{P}_D$. This is performed by counting the number of occurrences for which $\M>\cos(\theta_T)$ when $y(t)=n(t)$ and when $y(t)=s(t)+n(t)$, respectively. Unless stated otherwise, we determine the detection threshold based on a target $P_{\mathrm{fa}}=10^{-4}$, i.e., a CFAR detector. For efficiency, the sample buffer $y(t)$ and the reference signal $s(t)$ are downscaled baseband converted. As a byproduct, this verifies that our analysis above for real signals holds also for complex ones after a factor adjustment.

\subsection{Simulations}\label{sec:simulation}

The numerical simulations include transmission of a linear frequency modulation (LFM) chirp signal. The duration of the signal is set for $T_s=50$~msec, and its bandwidth varies with the considered $N$. Compliance with the system model, apart from the ambient noise, no channel distortion is used. The effect of the channel on performance is shown for the sea experiment discussed further below.

\begin{figure}[t]
\centering
\includegraphics[width=4.0in]{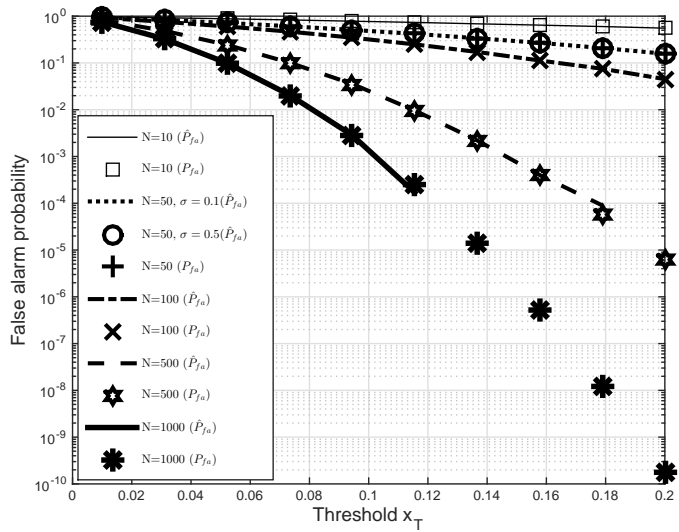}
\caption{Probability of false alarm as a function of detection Threshold.}
\label{f:3}
\end{figure} 
In Fig.~\ref{f:3}, results for the probability of false alarm are shown. Good match between the analysis ($p_{fa}$) and the empirical ($\hat{p}_{fa}$) results is observed. The results show the strong dependency between threshold $\theta_T$ from (\ref{e:pfa2}) and the compression ratio $N$. That is, for the same target probability of false alarm the threshold level dramatically decreases as $N$ increases. Fig.~\ref{f:3} also shows results of $\hat{P}_{\mathrm{fa}}$ for two $s$ / $\sigma$ ratios. As expected, the probability of false alarm does not depend on the SNR, i.e., the NMF detector is indeed a CFAR test.

\begin{figure}[t]
\centering
\includegraphics[width=4.0in]{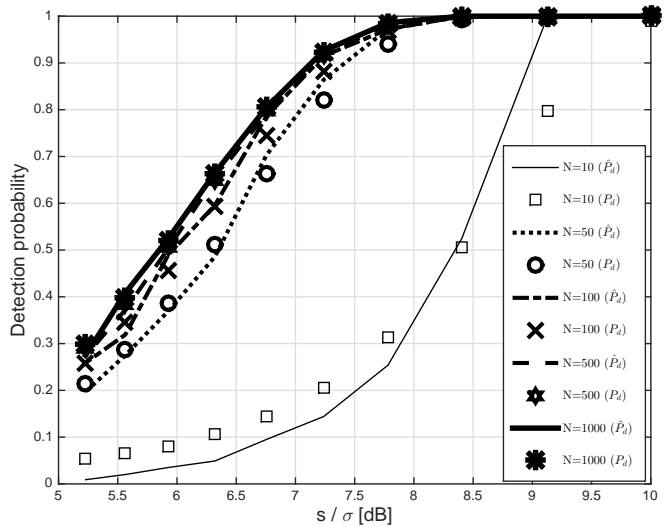}
\caption{Probability of detection as a function of $s$ / $\sigma$. Target $P_{\mathrm{fa}}=10^{-4}$.}
\label{f:4}
\end{figure} 
In Fig.~\ref{f:4}, approximation (\ref{e:pd2}) is verified for several compression ratios $N$. Analysis $p_{d}$ and the empirical $\hat{p}_{d}$ results are given. Results are shown as a function of $s$ / $\sigma$. One can observe that $P_D$ increases with $N$. This is because of the dependency of threshold $x_T$ in $N$. However, for high levels of $N$, $P_D$ saturates. Fig.~\ref{f:4} shows that for small values of $N$, there is only a rough match between the analytic approximation $P_D$ and the empirical measurement $\hat{P}_D$. However, for higher values of $N$ and starting from $N=50$, an excellent match is observed.

\begin{figure}[t]
\centering
\subfloat[\label{f:5a}]{\includegraphics[width=0.5\textwidth]{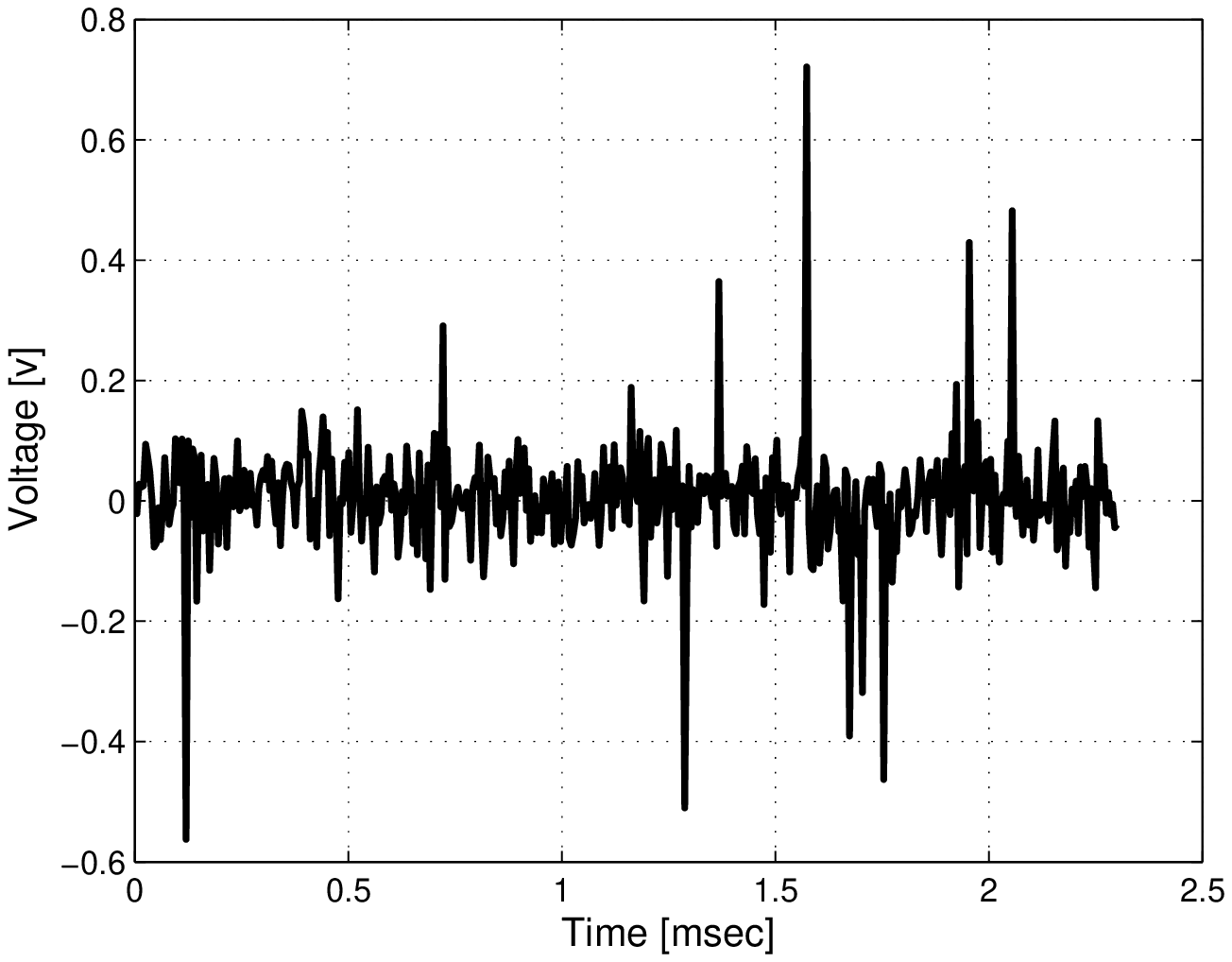}}
\subfloat[\label{f:5b}]{\includegraphics[width=0.5\textwidth]{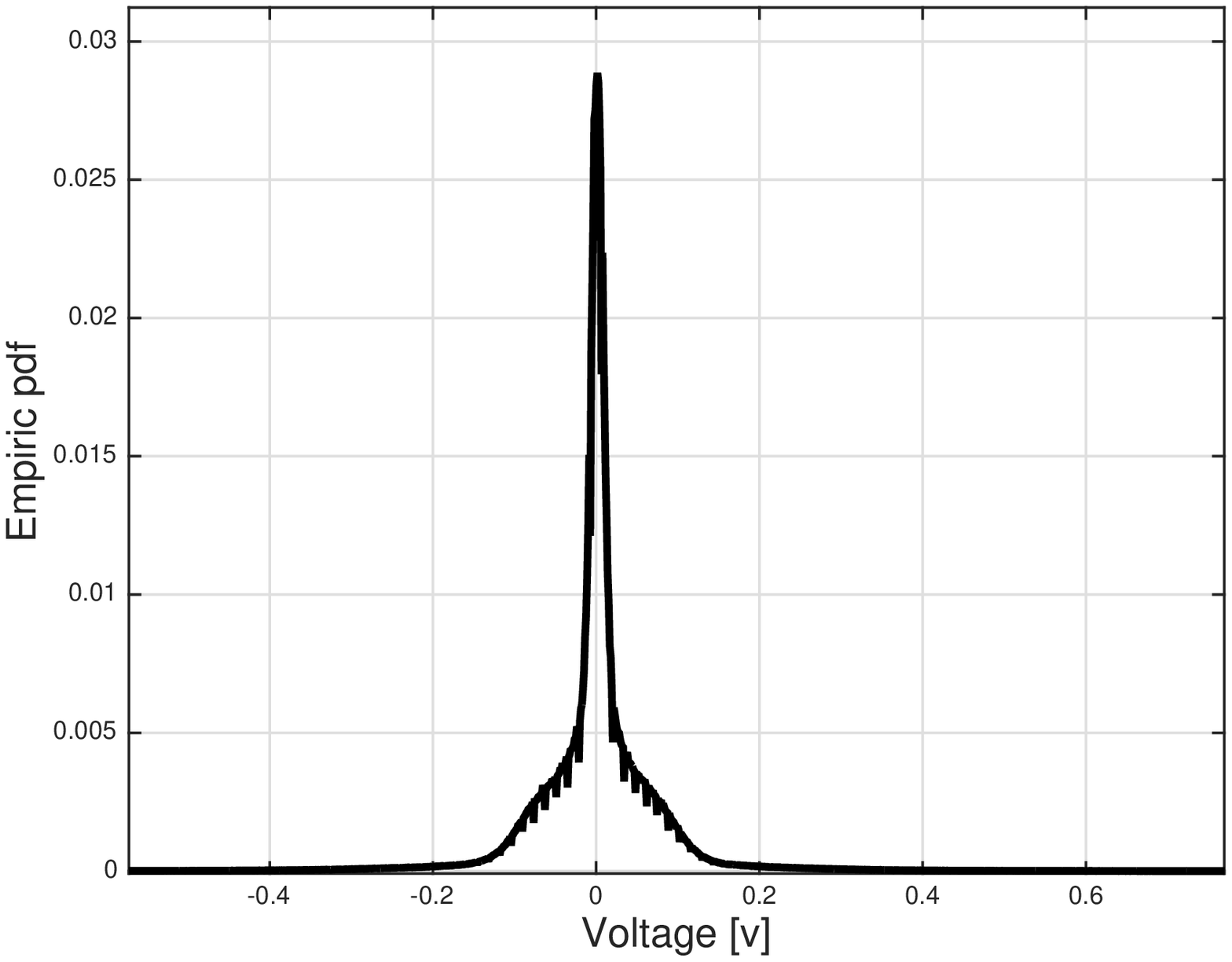}}
\caption{Recorded noise from sea experiment: (a) a single time-domain example, (b) empirical PDF.}
\end{figure}
\begin{figure}[t]
\centering
\subfloat[\label{f:6a}]{\includegraphics[width=0.5\textwidth]{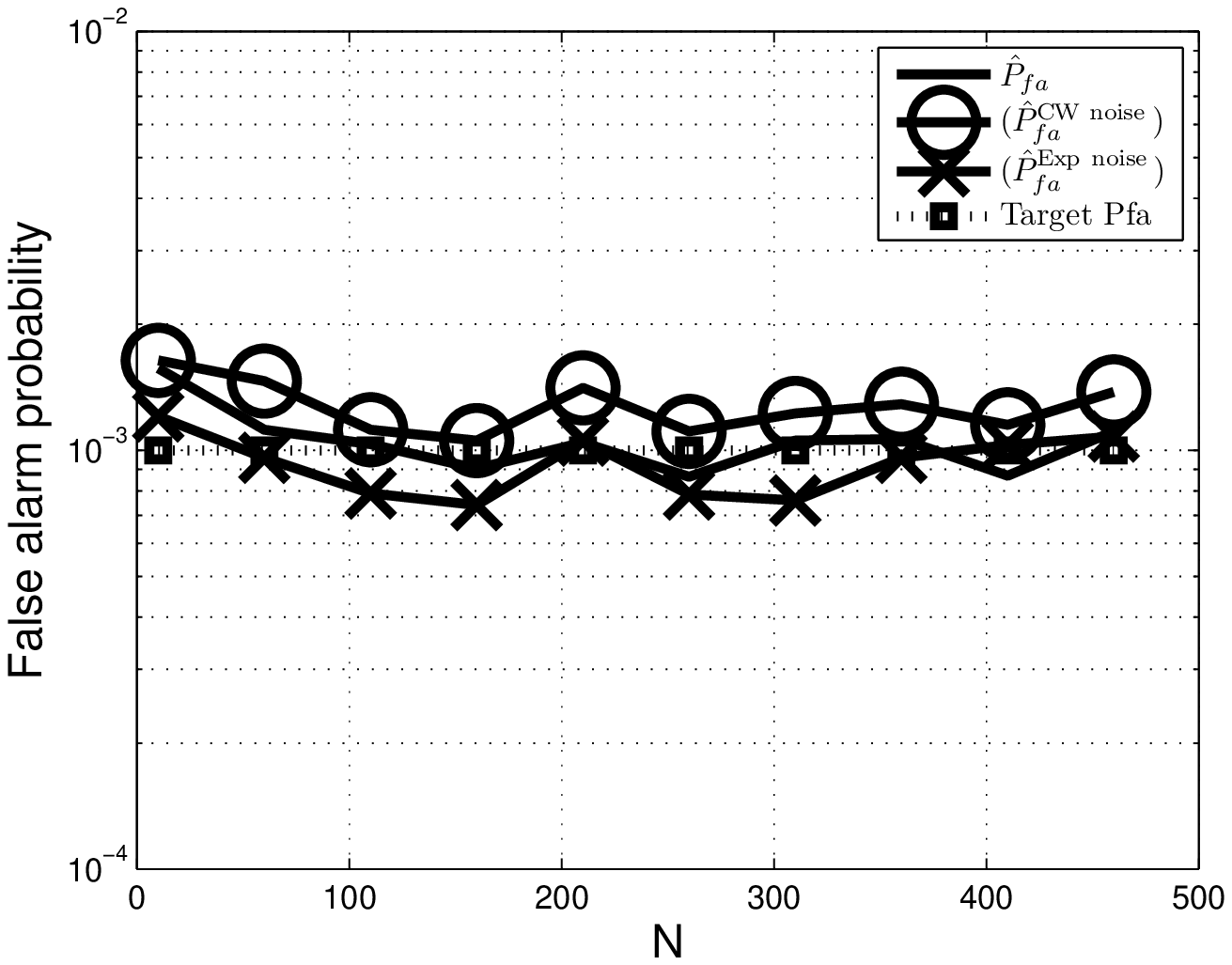}}
\subfloat[\label{f:6b}]{\includegraphics[width=0.5\textwidth]{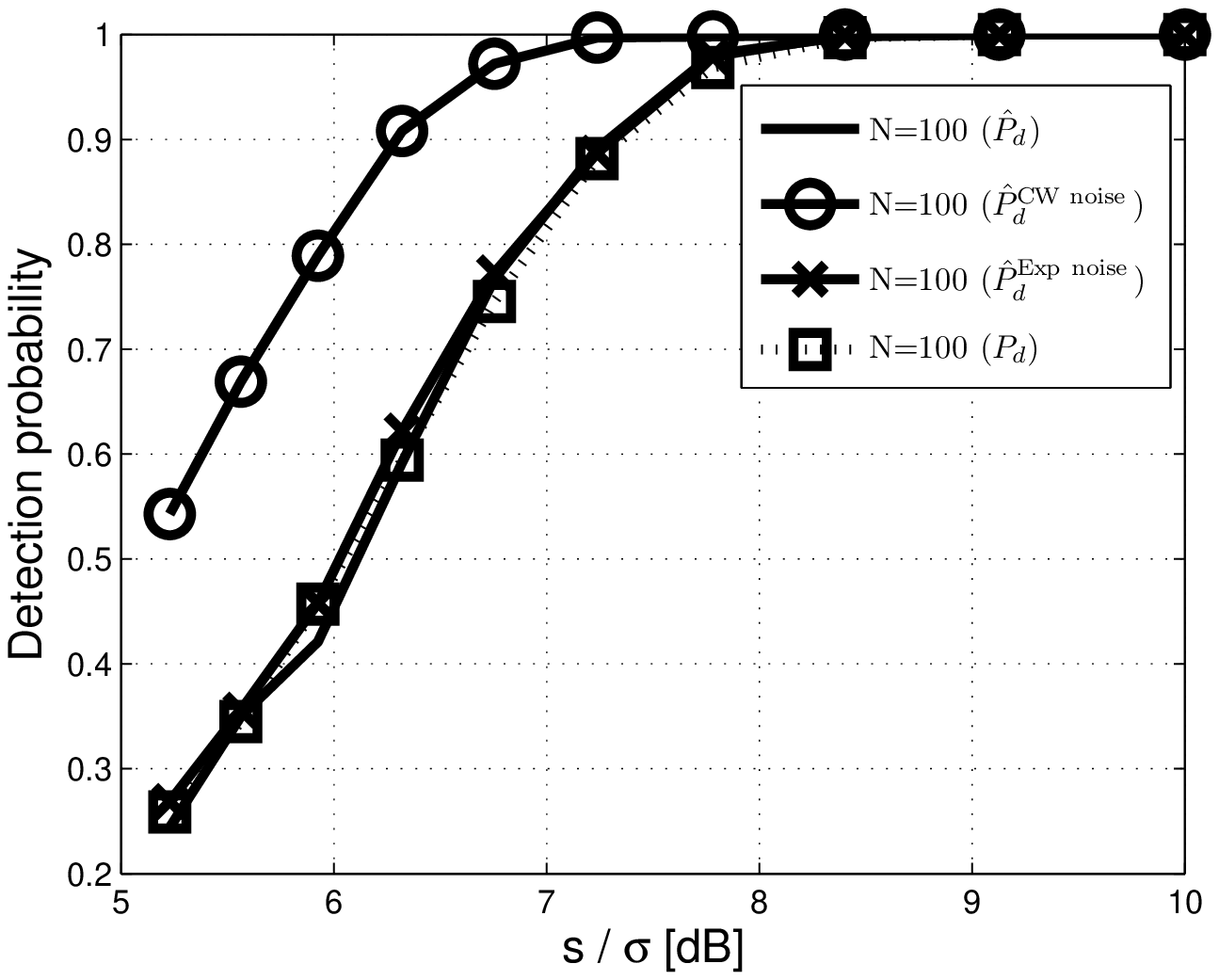}}
\caption{Detector performance: (a) Probability of false alarm as a function of $N$: effect of mismatch in noise model, (b) Probability of detection as a function of $s$ / $\sigma$: effect of mismatch in noise model. Target $P_{\mathrm{fa}}=10^{-4}$.}
\end{figure}

Next, the effect of a mismatch in the noise model is considered. Results for three different additive noise models are shown. Together with the i.i.d. zero-mean Gaussian noise ($\hat{P}_{d}$), the considered noise cased include a strong single carrier interference at the centre frequency ($\hat{P}_{d}^{\mathrm{CW \ noise}}$), and an ambient noise recorded during the sea experiment at different time windows ($\hat{P}_{d^{\mathrm{Exp \ noise}}}$). An example of the noise recorded in the experiment is shown in Fig.~\ref{f:5a}, and the empiric pdf evaluated from all noise instances recoded during the experiment is shown in Figure~\ref{f:5b}. One can observe the strong random transients in the experiment noise resulting wideband interferences, and the noise pdf appears to be similar to a Laplace Gaussian. The three noise components are normalized to test performance at different $s/\sigma$ ratios. Fig.~\ref{f:6a} shows results for the probability of false alarm, where $\hat{p}_{fa}$ represents results for i.i.d. Gaussian noise, $\hat{p}_{fa}^{\mathrm{CW \ noise}}$ represents results for noise with a CW interference, $\hat{p}_{fa}^{\mathrm{Exp \ noise}}$ represents results for noise recorded from the sea experiment, and Target $p_{fa}$ is the target false alarm probability. To reduce number of required simulation runs, the case considered is of a target false alarm of $10^{-3}$. Target false alarm probability is mostly achieved for the three noise models. Moreover, compliance with the results from Fig.~\ref{f:3}, this match is observed for small and large values of $N$. In Fig.~\ref{f:6b}, results are shown for the probability of detection for target false alarm of $10^{-4}$. Small differences are observed between the approximated analysis and the results for the recorded noise. A more significant effect is shown for the CW noise, where detection rate is better than the analysis. This is because, for a signal of large $N$, the NMF filters out narrowband interferences.

\begin{figure}[t]
\centering
\includegraphics[width=4.0in]{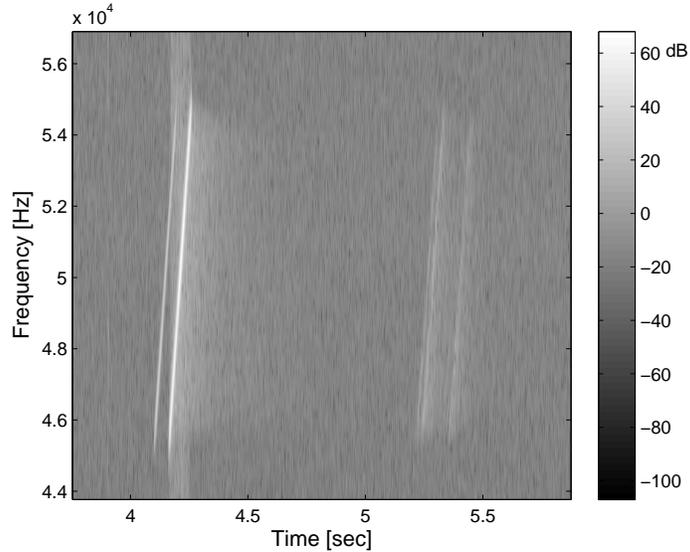}
\caption{Time-frequency response of sample buffer recorded during the sea experiment (depth roughly 900m).}
\label{f:7}
\end{figure} 
The results from the simulations verify the correctness of (\ref{e:pfa2}). Furthermore, for hydroacoustic signals where the common case is $N>50$, approximation (\ref{e:pd2}) predicts the probability of detection and thus the ROC. In addition, the results show that the analysis holds for an ambient noise consisting a single carrier interference and for the realistic case of noise recordings from a sea experiment.

\subsection{Sea Experiment}\label{sec:experiement}

To test the applicability of the system model and to demonstrate the accuracy of expressions (\ref{e:pfa2}) and (\ref{e:pd2}) in an actual sea channel, a sea experiment was conducted. The experimental setting was about 10~km off the shores of Haifa, Israel, at depth of roughly 900~m. The set included a surface vessel from which a transmitting projector and a receiving hydrophone have been deployed. The experiment included 600 hydroacoustic transmissions. Each transmission consisted of two linear frequency modulation chirp signals, spaced by a time gap of 100~msec, whose carrier frequency was 50~kHz, bandwidth 10~kHz, and whose duration varied between: 10msec, 50msec, and 100msec. The tested $N$ values were therefore 100, 500, and 1000. Several $s/\sigma$ values were tested by changing the amplification level of the transmitted signals. Transmissions were made at depth of 10~m, and signals were received at depth of 100~m. This depth difference allowed sufficient separation between the receptions of the direct path, surface path, and bottom path, while allowing the use of a narrow voltage range to reduce quantisation errors.

\begin{figure}[t]
\centering
\includegraphics[width=4.0in]{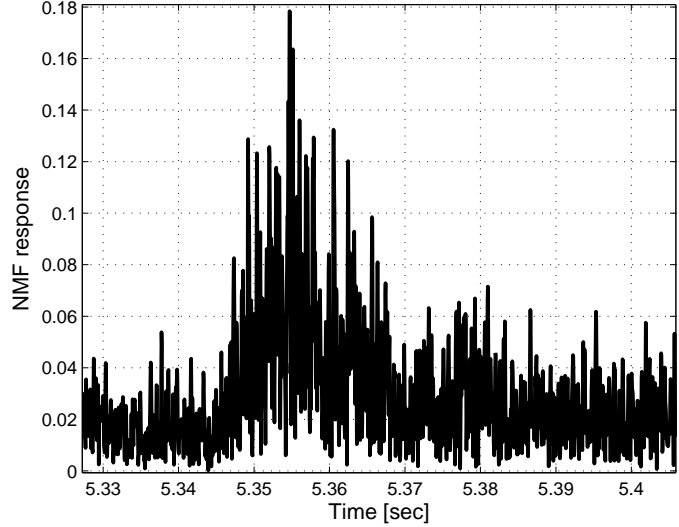}
\caption{Output of the NMF for the latest arrival of the sampled buffer from Fig.~\ref{f:8}.}
\label{f:8}
\end{figure} 
Following each transmission, a sampled buffer of 2~sec was collected from the channel. A time-frequency response of one of these sample buffers is shown in Fig.~\ref{f:7}. The reception of the two strong direct path is noticeable at time instances 4.1~sec and 5.3~sec, respectively. Strong surface reverberations are also observed. The much weaker signal reflected from the sea bottom can be seen at roughly 5.3~sec. For the latest arrival of the signals shown in Fig.~\ref{f:7}, the output of the NMF is shown in Fig.~\ref{f:8}. There exists a time difference of roughly 0.025~sec (or 38~m) between the first arrival and the last arrival. Since the transmitted signal is wideband (relative to the carrier frequency), this figure represents the length of the channel impulse response. For each detected signal, the $s/\sigma$ ratio was evaluated by estimating the arrival time of the received signal and measuring the signal energy and the noise level.

Detection of the two chirp signals was performed using the NMF detector with target $P_{\mathrm{fa}}=10^{-4}$ and detection threshold (\ref{e:pfa2}). Accurate detection was verified by comparing the time difference of arrival of each two local maxima of the NMF response with the expected time gap between the two transmitted chirp signals (100~msec). On the other hand, miss detection was declared when the output of the NMF did not exceed threshold. False alarm was determined for cases when the NMF response exceeded the threshold at wrong timing.

The decoding of the sea experiment achieved no false alarm. Considering the time window used for the sample buffer, this outcome corresponds to zero false alarms for roughly 2500 trials. The results for the detection rates are given in Table~\ref{t:ExpPdRes} alongside the predicted approximation (\ref{e:pd2}). Since for high values of $N$ the probability of detection changes little with $N$ (see Fig.~\ref{f:4}), results for $N\geq500$ are accumulated. In addition, for clarify, the measured $s/\sigma$ levels are quantized. The results in Table~\ref{t:ExpPdRes} show that, compliance with the analysis, no miss detection was found at $s/\sigma$ levels above 15~dB. However, for lower $s$ / $\sigma$ ratios, detection performances are below the expected level. This is explained by the effect of the non-linear channel (especially the Doppler shift phenomena and non-resolved multipath), which distorts the received signal and thus reduces the output of the NMF. Nonetheless, the performance gap is minor, and the results of the sea experiment mostly agree with the analysis.
\begin{table}
\begin{center}
\begin{tabular}{ | P{1.5cm} ' P{1.5cm} ' P{3.5cm} ' P{3.5cm} |}
    \Thickhline
    \textbf{$N$} & \textbf{$s/\sigma$ [dB]} & \textbf{Detection Rate (experiment)} &
     \textbf{Probability of Detection (analysis)}\\ \hline
     \Thickhline
    \multirow{5}{*}{100} & $\leq0$ & 0/57 (0)& 0.0022\\\hhline{~---}
    & 5 & 3/17 (0.17)& 0.21\\\hhline{~---}
    & 10 & 29/32 (0.9)& 1\\\hhline{~---}
    & 15 & 26/26 (1)& 1\\\hhline{~---}
    & $\geq$20 & 68/68 (1)& 1\\\hline
    \Thickhline
    \multirow{5}{*}{$\geq$500} & $\leq$0 & 1/51 (0.01)& 0.0021\\\hhline{~---}
    & 5 & 5/22 (0.22)& 0.235\\\hhline{~---}
    & 10 & 35/38 (0.92)& 1\\\hhline{~---}
    & 15 & 37/37 (1)& 1\\\hhline{~---}
    & $\geq$20 & 52/52 (1)& 1\\\hline
    \Thickhline
\end{tabular}
\end{center}
\caption{Results from Sea Experiment.}
    \label{t:ExpPdRes}
\end{table}

\section{Conclusions}\label{sec:Conclusions}
This paper focused on detection of hydroacoustic signals, where the time-bandwidth product, $N$, is large. The goal was to develop a computationaly efficient method for the determination of the detection threshold of the normalized matched filter (NMF). This detector is a CFAR scheme used when the noise covariance matrix is fast time-varying and is hard to estimate. The probability distribution and the moments of the NMF were derived. Then, both the exact finite-$N$ distribution (\ref{e:pd_distribution3}) and the large-$N$ limit (\ref{e:pd_distribution4}) were studied. For the case of large $N$ value, computational efficient expressions for the probability of false-alarm (\ref{e:pfa2}) and approximation for the probability of detection (\ref{e:pd2}) were developed. Using this analysis, a practical scheme was provided for calculating the receiver operating characteristic (ROC). Numerical simulations showed that the developed expressions are extremely accurate. Furthermore, the performance of the NMF detector was demonstrated in a sea experiment conducted at depth of 900~m off the coast of Haifa, Israel. The result of this work may serve as the basis for using the NMF as a practical detector for hydroacoustic signals.

%

%
\appendix

\section{Second Moment of a NMF}\label{sec:appendix}

In this section expression for the second moment of the NMF (\ref{e:NMF}) are developed for the case of $y(t)=n(t)$. For a sampled noise signal with a sampling period $\Delta$ and number of samples $N$,
\begin{equation}
\M^2=\frac{\left(\sum\limits_{k=1}^{N}n_ks_k\Delta\right)^2}{\left(\sum\limits_{k=1}^{N}n_k^2\Delta\right)\cdot\left(\sum\limits_{k=1}^{N}s_k^2\Delta\right)}\;.
\label{e:var_sample}
\end{equation}
Denote $\tilde{n}_k=\frac{n_k}{\sigma}$, $\tilde{s}_k=s_k\sqrt{\frac{\Delta}{E_s}}$, where $\sigma^2$ is the variance of $n(t)$ and $E_s$ is the energy of $s(t)$. Then,
\begin{equation}
\M^2=\frac{\left(\sum\limits_{k=1}^{N}\tilde{n}_k\tilde{s}_k\Delta\right)^2\sigma^2\frac{E_s}{\Delta}}{\left(\sum\limits_{k=1}^{N}\tilde{n}_k^2\Delta\right)\cdot\left(\sum\limits_{k=1}^{N}\tilde{s}_k^2\Delta\right)\sigma^2\frac{E}{\Delta}}\;.
\label{e:var_sample2}
\end{equation}
Clearly, $E\left[\M^2\right]$ does not depend on $\sigma$ or $E_s$. In the following, $s_k$ and $n_k$ are therefore refer to as normalized variables. The second moment of the sampled $\M$ is
\begin{equation}
E\left[\M^2\right]=E\left[\frac{\sum\limits_{k,l}^{N}s_ks_l\cdot n_kn_l}{\left(\sum\limits_{m=1}^{N}n_m^2\right)}\right]\;.
\label{e:var_sample3}
\end{equation}
To simplify (\ref{e:var_sample3}), one can use the connection
\begin{equation}
\frac{1}{X}=\int_{0}^{\infty}e^{-\lambda X}d\lambda\;,
\label{e:con}
\end{equation}
such that
\begin{equation}
E\left[\M^2\right]=\sum\limits_{k,l}^{N}s_ks_l\cdot E\left[\int_{0}^{\infty}n_kn_le^{-\lambda\sum\limits_{m}n_m^2}d\lambda\right]\;.
\label{e:var_sample4}
\end{equation}
Since $n_k$ is Gaussian, so is the integral in (\ref{e:var_sample4}) and
\begin{equation}
E\left[\M^2\right]=\sum\limits_{k}^{N}s_k^2\cdot E\left[\int_{0}^{\infty}n_k^2e^{-\lambda\sum\limits_{m}n_m^2}d\lambda\right]\;.
\label{e:var_sample5}
\end{equation}
Consider $N=1$. Here,
\begin{equation}
E\left[\M^2\right]=\int_{-\infty}^{\infty}\frac{dn}{\sqrt{2\pi}}\int_{0}^{\infty}n^2e^{-n^2\left(\lambda+\frac{1}{2}\right)}d\lambda=
\frac{\Gamma\left(\frac{3}{2}\right)}{\sqrt{2\pi}}\int_{\frac{1}{2}}^{\infty}\frac{da}{a^{\frac{3}{2}}}=1\;,
\label{e:var_example}
\end{equation}
where $a\def\lambda+\frac{1}{2}$ is used. The result in (\ref{e:var_example}) is a good sanity check since for the case of a single sample, the variance of the NMF is 1. For a general $N$,
\begin{equation}
E\left[\M^2\right]=\sum\limits_{k}s_k^2\cdot\frac{1}{\left(2\pi\right)^{\frac{N}{2}}}\Gamma\left(\frac{3}{2}\right)\pi^{\frac{N-1}{2}}\int_{\frac{1}{2}}^{\infty}\frac{da}{a^{\frac{3}{2}}a^{\frac{N-1}{2}}}=
\frac{\sqrt{\pi}}{2^{\frac{N}{2}}}\cdot\frac{1}{N}\pi^{-\frac{1}{2}}2^{\frac{N}{2}}=\frac{1}{N}\;.
\label{e:var_sample6}
\end{equation}
By (\ref{e:var_sample6}), the variance of $\M$ for the case of noise-only signal is inverse proportional to $N$.

\end{document}